\documentclass[%
 reprint,
preprintnumbers,
 amsmath,amssymb,
 aps,
]{revtex4-1}
\usepackage{cancel}
\usepackage{xcolor}

\usepackage[normalem]{ulem}

\usepackage{multirow}
\usepackage{graphicx}
\usepackage{dcolumn}
\usepackage{bm}
\usepackage{tikz}
\usepackage[compat=1.0.0]{tikz-feynman}

\begin{document}

\preprint{FERMILAB-PUB-23-363-T-V}

\title{New limits on $W_R$ from meson decays}

\author{Gustavo F. S. Alves}
\email{gustavo.figueiredo.alves@usp.br}
\affiliation{%
 Instituto de F\'isica, Universidade de S\~ao Paulo, C.P. 66.318, 05315-970 S\~ao Paulo, Brazil
}%
\affiliation{%
 Particle Theory Department, Fermilab, P.O. Box 500, Batavia, IL 60510, USA
}%
\author{Chee Sheng Fong}%
 \email{sheng.fong@ufabc.edu.br}
 \affiliation{Centro de Ciências Naturais e Humanas\\
Universidade Federal do ABC, 09.210-170, Santo André, Brazil}
 \author{Luighi P.  S. Leal}%
 \email{luighi.leal@usp.br}
 \author{Renata Zukanovich Funchal}%
 \email{zukanov@if.usp.br }
\affiliation{%
 Instituto de F\'isica, Universidade de S\~ao Paulo, C.P. 66.318, 05315-970 S\~ao Paulo, Brazil
}

\date{\today}

\begin{abstract}
In this letter we show that  pseudoscalar meson leptonic decay data can be used 
to set stringent limits on the mass $m_{W_R}$ of a right-handed vector
boson, such as the one that appears in left-right symmetric models. We have shown that for a heavy neutrino with a mass $m_N$ in the range $50<m_N/{\rm MeV} <1900$  one can constraint $m_{W_R} \lesssim (4-19)$ TeV at 90\% CL. 
This provides the most stringent experimental limits on the $W_R$ mass to date.

\end{abstract}

\maketitle

\textbf{\textit{Introduction.}}---
The weak interaction and its left-chiral nature has been connected
since its very inception  with neutrinos. 
On the one hand, except for gravity, 
neutrinos only interact weakly. On the 
other hand, all  neutrinos we have ever observed are left-chiral fermions ($\nu_L$). 
Furthermore, $\beta$-decays lead to 
the understanding that, at low energy,
weak interactions are governed by a universal constant, $G_F \sim 1/\Lambda^2 \sim 10^{-5}$ GeV$^{-2}$, the Fermi constant. This, retrospectively, was the first indication of the need for a mediator with mass $\Lambda \sim {\cal O}(100 \; \rm GeV)$, for couplings of $\mathcal{O}(1)$.
So neutrino properties 
were at the core of building the Standard Model (SM) of electroweak interactions as a left-chiral gauge theory.

Neutrino oscillation experiments 
have provided to us in the last half of century compelling evidences that neutrinos have (tiny) masses 
and undergo flavor mixing~\cite{Cleveland:1997zz, Kamiokande:1996qmi, Kamiokande-II:1992hns, SNO:2001kpb, GNO:2000avz, SAGE:1999uje, Super-Kamiokande:1998qwk, Super-Kamiokande:2001ljr, Super-Kamiokande:1998kpq, Super-Kamiokande:2006yyp, Becker-Szendy:1992ory, Allison:1996yb, Kamiokande:1994sgx, KamLAND:2002uet, KamLAND:2004mhv, DoubleChooz:2011ymz, DayaBay:2012fng, RENO:2012mkc, T2K:2011ypd, MINOS:2011amj, K2K:2006yov, K2K:2004iot}. 
We may need right-chiral 
neutrino fields $N$  to explain 
neutrino masses and mixings, however, 
these states are uncharged under $SU(2)_L \times U(1)_Y$, the SM symmetry group. This is why we sometimes refer to left(right)-chiral neutrinos as active (sterile). 

We do  not know if there are right-chiral weak charged currents in nature. If they 
exist  right-chiral neutrinos would be active under them. So low energy weak decays involving neutrinos can be used to test their effective strength $G_F^\prime$ and probe the corresponding mass scale of the new mediator $W_R$. 

We will focus here on two body pseudoscalar meson decays $M \to \ell \, N$, where  $M=\pi, K$ and $ D$, $\ell=e,\mu$ and $N$ a right-handed neutrino in the MeV--GeV mass range. In extensions of the SM with right-chiral currents and right-chiral neutrinos, the decay rate $\Gamma(M \to \ell \, N)$ can have two competing contributions \footnote{We have ignored the interference term as it does not play a role when either mixing or the right-handed current dominates.} 
\begin{equation}
\Gamma(M \to \ell \, N) =( G_F^2 \vert U_{\ell N}\vert^2 
+ G_F^{\prime 2} ) \; f(m_M,m_\ell,m_N) \, ,
\end{equation}
the first one mediated by $W_L$, the 
SM weak vector boson, and depends on the
active-sterile mixing $U_{\ell N}$ between $\nu_{\ell L}$ and $N$, the second mediated by $W_R$. Here $f(m_M, m_\ell,m_N)$ is a function 
that will depend on the meson mass $m_M$, 
charged lepton mass $m_\ell$ and right-handed neutrino mass $m_N$.  The two low energy effective couplings are related by 
\begin{equation}
\left(\frac{G_{F}'}{G_{F}}\right)^2 \equiv \left(\frac{m_{W_L}\, g_R}{m_{W_R}\, g_L}\right)^{4}\sim 7\times 10^{-8}\left( \frac{5 \, \rm TeV}{m_{W_R}}\right)^4
\kappa^4
\,,
\label{eq:GF}
\end{equation}
where $m_{W_L}$ and $g_L$  ($m_{W_R}$ and $g_R$) are the mass and coupling 
constant associated with the SM (new) weak interaction, and $\kappa \equiv g_R/g_L$. If $\vert U_{\ell N}\vert^2 \gg (G_F^{\prime}/G_F)^2$ the mixing contribution prevails and 
 one can use meson decays to constrain  the active-sterile mixing, 
 as has been done by several 
 authors~\cite{Bryman:2019bjg, Ballett:2019bgd, Barouki:2022bkt, T2K:2019jwa, PiENu:2015pkq, CHARM:1985nku, NA62:2020mcv, BEBCWA66:1986err}. However, 
 if $\vert U_{\ell N}\vert^2 \ll (G_F^{\prime}/G_F)^2$ the right current contribution dominates and  one can 
 use these decays instead to constrain 
 $m_{W_R}$. 
 The best limits on active-sterile mixing are on $U_{eN}$. While 
 they depend on $m_N$, in the mass 
 range of interest the maximum value allowed  by data is around  
 $\vert U_{eN}\vert^2 \sim 10^{-7}-10^{-9}$, 
 which imply  meson decay experiments can have a  sensitivity to 
 $m_{W_R} \sim (5-15)$ TeV. 
 Note that this sensitivity is comparable and even surpasses the best limits on $m_{W_R}$ we currently  have  from  the LHC experiments $m_{W_R}<6.4$ TeV~\cite{ATLAS:2023cjo, CMS:2021dzb}.
 
 In this letter we reanalyze the results  
 from a number of  low energy meson  decay experiments  under the 
 assumption of right-chiral current dominance, 
 a situation not considered before in the literature, and that, theoretically, may manifest in left-right symmetric models (LRSM)~\cite{Pati:1974yy, Mohapatra:1974gc, Mohapatra:1979ia, Mohapatra:1980yp, Senjanovic:1975rk, Deshpande:1990ip}, to derive the best experimental limits to date on $m_{W_R}$.
 
 \textbf{\textit{Left-right Symmetric Models.}}---
LRSM remain arguably one of the simplest and best motivated extensions of the SM. 
Being characterized by the gauge group $SU(2)_L\times SU(2)_R \times U(1)$ 
and an additional discrete LR symmetry~\cite{Nemevsek:2012iq, Senjanovic:2016vxw}, 
they forecast the existence of two new gauge bosons: a neutral $Z_R$ and a charged $W_R$.  Fermions are LR symmetric, i.e., $q_{L,R}=(u \; d)^T_{L,R}$ and $\ell_{L,R}=(\nu \; e)^T_{L,R}$, and 
the $SU(2)_{L,R}$ associated gauge couplings $g_L$ and $g_R$ can either be equal or not, depending on the discrete LR symmetry breaking scale~\cite{Chang:1984uy}.

Neutrino masses are natural to these 
models as three right-chiral neutrinos $N \equiv \nu_R$ have to be introduced to complete the $SU(2)_R$ lepton doublets. Furthermore, the light neutrino masses, can be made 
small, via the contributions of type-I and type-II seesaw mechanisms~\cite{Minkowski:1977sc, Mohapatra:1979ia,Glashow:1979nm,Gell-Mann:1979vob}, that is
\begin{eqnarray}
m_{\nu} & = & m_{\textrm{I}}+m_{\textrm{II}}.
\label{eq:mnu}
\end{eqnarray}
Note that in a type I dominant scenario, $m_\nu \sim m_{\textrm{I}} \sim \vert U_{\ell N}\vert^2 m_N$, so to fulfill our requirement on subdominant active-sterile mixing,  we need 
\begin{eqnarray}
m_{\nu} & < & 7\times10^{-2}\,\textrm{eV}\left(\frac{m_{N}}{1\,\textrm{MeV}}\right)\left(\frac{5\,\textrm{TeV}}{m_{W_R}}\right)^{4} \kappa^4\,,
\label{eq:typeI}
\end{eqnarray}
which can, in principle,  hold for 
$m_N$ in the MeV-GeV range. In a type II dominant scenario $m_\nu \sim m_\textrm{II}$ since $|U_{\ell N}|^2 \sim m_I/m_N$ the mixture contribution can always be made small.  
We will disregard the active-sterile mixing contribution from now on by setting $U_{\ell N}=0$.

The relevant part of the model Lagrangian 
for our study is 
\begin{eqnarray}
{\cal L^{\rm cc}_{\rm R}} = - \frac{g_R}{\sqrt{2}}[\overline{N} \, U_{RR}^\dagger \, \cancel{W_R}  \, E_R + 
\overline{D}_R \, V_R^\dagger \, \cancel{W_R} \, U_R] + \rm h.c. \, ,
\label{eq:lagrangian}    
\end{eqnarray}
where the right-chiral fermion fields are grouped as $N=(N_1\, N_2 \, N_3)^T$ for  neutrinos, $E_R=(e_R \, \mu_R \, \tau_R)^T$ for charged leptons, $D_R=(d_R \, s_R \, b_R)^T$ for down- and 
$U_R=(u_R\, c_R \, t_R)^T$ for up-quarks.
 The Lagrangian is given in the mass basis so $U_{RR}$ and $V_R$ are unitary mixing matrices. We will set $V_R=V_{\rm CKM}$, the same CKM mixing matrix of the SM. As shown in ref.~\cite{Senjanovic:2014pva}, this relation holds to a very high degree in the minimal left-right model (and is exact for real bidoublet vacuum expectation values) despite the left-right symmetry being badly broken. Furthermore, to make our analysis as model-independent as possible, we will also assume all right-handed neutrinos to have the same mass $m_N$ such that $U_{RR}$ drops out from the calculations. If they are not degenerate in mass, the calculations will involve $U_{RR}$ and there might be new decay channels from heavy $N$ to lighter $N$, we will leave this model-dependent analysis for future work.

We can safely disregard mixing between  $W_R$ and $W_L$~\footnote{Mixing between $W_L$ and $W_R$ can be generated by electroweak  corrections~\cite{Branco:1978bz}, the most significant involving top and bottom quarks and so being of the order of $(\alpha/4\pi) (m_t \, m_b)/m^2_{W_R} \ll (G_F^{\prime}/G_F)$.}. Finally, notice that by setting a limit on $m_{W_R}$, we are also indirectly constraining the mass of $Z_R$ from the mass relation after breaking the LR symmetry.

\begin{figure}[!h]
    \centering
    \begin{tikzpicture}
      \begin{feynman}
        \vertex (a1) {\( \)};
               \vertex [below right =of a1] (a2);
        \vertex [right=of a2] (a3);
        \vertex [above right=of a3] (a4) {\(N\)};
        \vertex [below right=of a3] (a5) {\(l^{+}\)};
    
        \vertex[below=6em of a1] (b1) {\( \)};
         
        \diagram*{
          (a1) -- [fermion] (a2),
          (a2) -- [boson, edge label'=\(W_R^{+}\)] (a3),
          (a3) -- [fermion] (a4),
          (a3) -- [anti fermion] (a5),
    
          (b1) -- [anti fermion] (a2),
        };

       \draw[fill=black] (0,-1) ellipse (.2cm and 1cm) node [pos=0.5, left] {\(M^{+}\)};
      \end{feynman}
    \end{tikzpicture}
    \caption{Production of a right-handed neutrino $N$ by the meson $M$ decay mediated by the right-handed current.}
    \label{fig:N_production}
\end{figure}
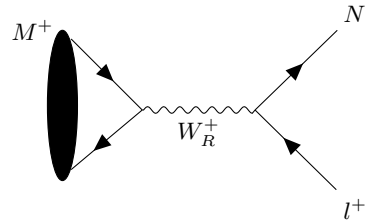

\textbf{\textit{Right-handed neutrino searches.}}---
The primary production mechanism 
for $N$ in accelerators are two-body 
pseudoscalar meson decays.
In the limit where the active-sterile 
mixing is suppressed, this is accomplished via the tree-level process mediated by $W_R$  depicted in Fig.~\ref{fig:N_production}. Because 
all right-handed neutrinos of the model are degenerate in mass there is no mixing suppression in the leptonic vertex. So 
the rate of this process is like the one for  $\Gamma(M \to \ell_L \, \nu_{\ell L})$ in the SM, except for the exchange $G_F \to G_F^{\prime}$ and the correction to the 
matrix element and phase-space due to a non-negligible $m_N$. Similarly, for detection, only channels mediated via the charged right-handed current must be taken into account. There are three types of such searches: visible (with hadrons in the final state), invisible  and meson decay ratios.

\textbf{\textit{Visible Searches.}}---The first class of experiments we will discuss look for visible signals from 
$N \to \ell^\pm \pi^\mp$ decay 
in the detector.  

We start with the Tokai-to-Kamioka (T2K) 
experiment, a neutrino oscillation experiment in Japan~\cite{T2K:2011qtm}. T2K beam 
is produced mainly by $\pi$ and $K$ 
decays which result from the collision of 30 GeV protons on a graphite target. These mesons are focused and their charge is selected by magnetic horns before they decay in flight producing neutrinos.
We say they operate in the neutrino (antineutrino) mode for positive (negative) charged meson selection.

The collaboration has used data 
collected by their off-axis near detector, ND280, to look for $N$
visible decays. They assume $N$ is  
produced and decay via active-sterile mixing. Their analysis correspond 
to an exposure to $12.34 \times 10^{20}$ protons on target  (POT) in the neutrino  mode and $6.29 \times 10^{20}$ POT in the antineutrino mode. The ND280 is 
a detector located 280 m from the proton target and having 
three time projection chambers (TPC) as their  central tracker surrounded by a calorimeter and a  muon detector~\cite{T2KND280TPC:2010nnd}.
The main active volume is the $6.3$ m$^3$ for gas TPC.
They have considered the production  
modes $K^\pm \to \ell^\pm \, N$, with $N$ having a life-time sufficiently long ($\tau \gg 1 \, \mu $s)  so it can 
reach ND280, where it will subsequently  decay in one of the following modes $N\to \ell^\pm \, \pi^\mp$, $N\to \ell^\pm \ell'^\mp \nu$, with $\ell,\ell'=e,\mu$.
The main background  they expected are from  neutrino coherent $\pi$ production in Ar ($\nu_\mu + {\rm Ar} \to \mu^- + \pi^+ + {\rm Ar}$), but they also consider  other neutrino interactions in and outside the gas TPC.
In Tab.~II of \cite{T2K:2019jwa} we can find the background they have estimated for each production and decay mode (typically $<1$ event), as well 
as the effect of Monte Carlo (MC) statistics, flux and detector systematics in their background calculation ($<0.5$ events). 
They have not observed any events in any 
of these modes in their data.

Only the two body production and decay modes  with the same flavor charged lepton can be due to $W_R$ under our assumptions: $K^\pm \to e ^\pm N, \; N \to e^\pm \pi^\mp$ for $m_\pi +m_e < m_N < m_K - m_e$ (four channels) and $K^\pm \to \mu ^\pm N, \; N \to \mu^\pm \pi^\mp$ for $m_\pi+m_\mu < m_N < m_K-m_\mu$ (four channels).

Limits complementary to T2K are provided by reanalysing the constraints from the Big European Bubble Chamber (BEBC) \cite{BEBCWA66:1986err} on heavy neutral leptons \cite{WA66:1985mfx}. The neutrino beam is originated from a beam dump setup where a flux of $400\ \rm GeV$ protons hit a copper block thick enough to absorb the long lived mesons produced in the collision before they decay. As a result, the expected right-handed neutrino flux predominantly consists of prompt $D^\pm \to 
\ell^\pm \, N$ decays, enabling exploration of masses in the range $250 < m_N/\rm{MeV} \lesssim 1900$. The bubble chamber detector is positioned $406\ \rm m$ from the copper layer, followed by other two neutrino detectors, CDHSW and CHARM. Analysis of the data collected by the BEBC experiment has led to strong constraints on the mixing of heavy neutral leptons with muon and electron neutrinos \cite{WA66:1985mfx}. The detection channels considered  were the same as in the T2K experiment. The total amount of data corresponds to $\sim 2 \times 10^{18}$ POT. They 
have observed a single event in the 
$N \to \mu^+ \pi^-$ mode, which is consistent to their expected background 
of $0.6 \pm 0.2$ events.

Unfortunately we cannot profit from  
the CHARM experiment 
data~\cite{CHARM:1985nku} used to derive 
limits on active-sterile mixing 
produced by charm meson decays because they only consider the three body final states $N\to \ell \ell' \nu$, which cannot occur via $W_R$ for $U_{\ell N} =0$. 

\textbf{\textit{Invisible Searches.}}---
The second class are peak 
search experiments. These experiments
look for the existence of a heavy neutrino emitted in two body helicity suppressed meson decays $M^+ \to  e^+ \nu_e (N)$. 
The charged meson can decay either at rest or 
in flight, in both cases the signal 
$M^+ \to e^+ N$ is characterized  by a single final state positron, similar to the SM decay.
The idea is to search for a subdominant peak  in the $e^+$ spectrum~\cite{Shrock:1980vy,Shrock:1980ct} due to the presence of an invisible particle of mass $m_N$. The peak-search procedure measures the $M^+ \to e^+ N$ decay rate with respect to  $M^+  \to e^+ \nu_e$, the SM rate, as a function of $m_N$.
 This approach profits from  major cancellations of the residual  
 inefficiencies not fully accounted for in MC simulations but present in both 
 signal and normalization modes.
Further corrections can be accounted 
for by the signal selection acceptance. In the case we are 
considering these branching ratios 
are related by
\begin{equation}
{\cal B}(M^+ \to e^+ N) =     {\cal B}^{\rm SM}(M^+ \to e^+ \nu_e) \, \rho_e^{MN}\, \left(\frac{G_F^{\prime}}{G_F}\right)^2\,,
\label{eq:BR}
\end{equation}
where $x_e=(m_e/m_M)^2$,  $x_N=(m_N/m_M)^2$ with $M=\pi$ or $K$, and 
the corresponding kinematical factor is 
$\rho_e^{MN}=[x_e+x_N-(x_e-x_N)^2] \lambda^{1/2}(1,x_e^2,x_N^2)/[x_e(1-x_e)^2]$ with $\lambda(a,b,c)=a^2+b^2+c^2-2(ab+bc+ac)$.

We will focus here on experiments PIENU~\cite{PiENu:2015pkq} and NA62~\cite{NA62:2020mcv}. 
For $M_{W_R} \gtrsim 5$ TeV, $N$ has a lifetime $\tau_N \gg 1\ \mu$s so it
can be considered stable in these production-search experiments.
The PIENU detector at TRIUMF uses a 
secondary pion beam created by colliding 500 MeV protons into a beryllium  target. The  positively charged beam (84\% $\pi^+$, 14\% $\mu^+$ and 2\% $e^+$) of momentum 75 MeV is transported to the PIENU apparatus.
The $\pi^+$ are stopped in a 8 mm thick plastic scintillator and decay at rest. 
The  positrons, which are monocromatic ($E_{e^+}=69.8$ MeV), are measured in a spectrometer consisting of a large NaI (T$\ell$) crystal (48 cm long and 48 cm diameter)  surrounded by an array of pure 
CsI crystals.
They have collected  about $10^7$ $\pi^+ \to e^+ \nu_e$ events, which they used 
to look for $N$ production via active-sterile mixing $U_{eN}$ for $60< m_N/{\rm MeV} < 135$~\cite{PIENU:2017wbj}. Their main background is $\pi^+ \to \mu^+ \nu_\mu$ 
followed by $\mu^+ \to e^+ \nu_e \bar\nu_\mu$. They were able to 
suppress this background by applying  cuts on timing, energy and track information.  Their MC simulation was validated with an experimental study~\cite{PIENU:2017wbj}. 
Their background suppressed positron 
spectrum was fitted with a background 
and a signal component for 
$E_{e+} = 4 $ MeV to $E_{e+} = 56 $ MeV 
in order to search for additional peaks. 

The TINA detector~\cite{Britton:1992xv} is an older TRIUMF experiment 
very similar to PIENU. It 
has also looked for peaks in the 
the positron spectrum from pion  
decays at rest  but using  a total of 
$1.2 \times 10^{5}$ $e^+$ events.
It has lower sensitivity than PIENU, except in the lower part of the range $50<m_N/{\rm MeV}<130$ \footnote{A similar route was pursued by the SIN experiment \cite{PhysRevD.36.2624}, where they searched for peaks in the energy spectrum of the muons produced from pion decays. Their result probe a lower mass range  $1<m_N/{\rm MeV}<20$, but the sensitivity is comparatively lower than the bounds from decay ratios discussed in the next section.}.

The NA62 detector at CERN uses a secondary 
beam ( 70\% $\pi^+$, 23\% protons  and 6\% $K^+$) created by directing 400 GeV protons from the SPS onto a beryllium target. The central beam momentum is 75 GeV, with a momentum spread of 1\%.
Before entering the long fiducial decay volume of the detector $K^+$ are tagged by a Cherenkov counter and 
hadrons from $K^+$ upstream decays are absorbed by a steel collimator~\cite{NA62:2020mcv}. The 
momenta of charged particles produced by $K^+$ decays  are measured by a magnetic spectrometer. To maximize signal and avoid background, the $e^+$ track momentum is restricted to be in the  (5-30) GeV range and the reconstructed squared missing mass $m_{\rm miss}^2=(p_K-p_{e+})^2< 0.01$ GeV$^2$, where $p_K(p_{e+})$
is the kaon (positron) four-momentum.
Their available data corresponds to 
 $0.79 \times 10^6$ SPS spills recorded during 360 days of operation in 2017–2018, at a typical beam intensity of $2.2 \times 10^{12}$ protons per spill. 
 They have looked for $N$ produced by active-sterile mixing with a 
 lifetime exceeding 50 ns, and considering that after being  produced in  $K^+ \to e^+ N$ decays they would be 
 boosted by a Lorentz factor of ${{\cal O}(100)}$,  so their decay into SM particles in the 156 m long volume between the start of the fiducial volume and the last detector  can be neglected. 
 They have analysed data corresponding to $N_K=(3.52\pm 0.02)\times 10^{12}$
 kaon decays in the fiducial volume. 
They have investigated 264 mass hypotheses, $m_N$, with $144< m_N/{\rm MeV} < 462$. 
 Their dominant background is 
 due to $K^+ \to \mu^+ \nu_\mu$
followed by $\mu^+ \to e^+ \nu_e \bar \nu_\mu$. This is reduced by 
requiring compatibility between the $e^+$ and $K^+$ tracks.
Other backgrounds,  including $K^+ \to \mu^+ \nu_\mu$ with a misidentified muon, are negligible according to ref.~\cite{NA62:2020mcv}.

 \textbf{\textit{Meson Decay Ratios.}}---
The third class of searches investigate 
the effect of $N$ in the ratio of 
pseudoscalar meson leptonic decays to $e$ and $\mu$ final states~\cite{Shrock:1980ct}, constraining the ratio
\begin{equation}
R_{e/\mu}(M) = \frac{1+ R_{N/\nu_e}(M)}{1+R_{N/\nu_\mu}(M)} \; R^{\rm SM}_{e/\mu}(M)\, ,
\label{eq:remu}
\end{equation}
where $R_{e/\mu}^{\rm SM}(M)\equiv {\cal B}^{\rm SM}(M \to e \, \nu_e)/{\cal B}^{\rm SM}(M \to \mu \, \nu_\mu)$ and 
$R_{N/\nu_\ell}(M) \equiv {\cal B}(M \to \ell\,  N)/{\cal B}^{\rm SM}(M \to \ell \, \nu_\ell)$ with respect to the experimental values
$R^{\rm PDG}_{e/\mu}(\pi)= (1.2327\pm 0.0023) \times 10^{-4}$ and 
$R^{\rm PDG}_{e/\mu}(K)=(2.488\pm 0.009) \times 10^{-5}$.
Since the leading order radiative corrections do not depend on $m_N$~\cite{Marciano:1993sh,Bryman:2019bjg} we consider they are the same for $R_{e/\mu}$ (eq.~(\ref{eq:remu})) and $R_{e/\mu}^{\rm SM}$.
We will use here the SM predictions $R^{\rm SM}_{e/\mu}(\pi)= (1.2352\pm 0.0001) \times 10^{-4}$ and 
$R^{\rm SM}_{e/\mu}(K)= (2.477 \pm 0.001) \times 10^{-5}$~\cite{Cirigliano:2007xi}.
Note that in calculating $R_{N /\nu_\ell}$ we have to take into account which  decay channels will be available depending on $m_N$.

\textbf{\textit{Results.}}---
Our main results are presented in Fig.~\ref{fig:Bounds_MW},
where we show the exclusion in the plane $(m_N, m_{W_R})$ for $\kappa=1$.
The T2K bound for $140 < m_N/{\rm MeV}< 493$ was calculated using 
the public MC simulation of the expected signal after geometrical, kinematical and efficiency cuts. 
This is available as a table with the expected number of events in the detector per production and decay modes as a function of $m_N$ assuming 100\% selection
efficiency and $U_{\ell N}=1$~\cite{T2K:2019jwa}. 
We have  used this table to compute the expected  number of events as a function of $m_{W_R}$ and $m_N$ simply by  
re-scaling the relevant channels by $(G_F^{\prime}/G_F)^2$. 
The sensitivity to $m_{W_R}$ increases with $m_N$ until about 388 MeV (for larger $m_N$ the four channels involving the $\mu$ cannot contribute anymore), reaching $m_{W_R} \gtrsim 14$ TeV. However, it remains high up to 493 MeV, partially due to high flux and background suppression~\footnote{The background 
in the neutrino mode is about 5  times smaller for $N\to e^\mp \, \pi^\pm$ with respect to $N \to \mu^\mp \, \pi^\pm$.}. The BEBC limit was obtained as follows.
 The $N$ flux was inferred from the light neutrino flux, taking into account only the two-body decays of $D$ mesons. To that end we adapted the simulation provided in~\cite{Barouki:2022bkt} to include only the channels mediated by $W_R$  and re-scaling the number of events by $(G_F^{\prime}/G_F)^2$. 
We get $m_{W_R} \gtrsim (4-5)$ TeV. These are the best limits in the region $500<m_N/{\rm MeV}<2000$. Note, however, that there is a small (white) region for small $m_{W_R}$ which cannot be discarded by BEBC. There the $N$ flux is suppressed because most of these particles decay before reaching the detector.
 In both experiments the 
 exclusion region is incompatible with the expected background at 90\% CL.

In the case of the peak searches 
we have taken Fig.~5~\cite{PIENU:2017wbj},
Fig.~3b (curve A)~\cite{Britton:1992xv} and Fig.~5~\cite{NA62:2020mcv}, for  
PIENU, TINA and 
NA62, respectively, and calculated the 90\% CL exclusion using the conversion  
$\vert U_{eN}\vert^2 \to (G_F^{\prime}/G_F)^2$.
These searches limit $m_{W_R}<(4-19)$ 
TeV, depending on $m_N$. 
TINA gives the best limit for $50<m_N/{\rm MeV}\lesssim 60$, PIENU for  $60\lesssim m_N/{\rm MeV}\lesssim 130$ and NA62 
for $144<m_N/{\rm MeV}\lesssim 440$.

Finally, we have used 
\begin{align}
R_{e/\mu}(\pi)/R^{\rm SM}_{e/\mu}(\pi)&<  \overline{R}^{(\pi)}_{e/\mu}({\rm PDG})=1.0017\, ,\nonumber \\
R_{e/\mu}(K)/R^{\rm SM}_{e/\mu}(K)&< 
\overline{R}^{(K)}_{e/\mu}({\rm PDG}) =1.012 \, ,
\end{align}
where $\overline{R}^{(M)}_{e/\mu}({\rm PDG})\equiv (R^{\rm PDG}_{e/\mu}(M)+ 2\sigma)/R^{\rm SM}_{e/\mu}(M)$ 
to compute the meson decay ratio limits on Fig.~\ref{fig:Bounds_MW}. The  $\overline{R}^{(K)}_{e/\mu}({\rm PDG})  $
is  dominant in the gap 
between PIENU and NA62 $ m_N \sim 0.13\ \rm{GeV}$ and for $m_N \lesssim 0.05\ \rm{GeV}$ where it is the only bound applicable. In this mass range not shown in the plot, the constraint gets weaker starting from $m_{W_R} \sim 4$ TeV for $m_N \sim 0.05\ \rm{GeV}$ down to $m_{W_R} \sim 0.5\ \rm{TeV}$ for $m_N \sim 1\ \rm{MeV}$.

\begin{widetext}
    \begin{center}
        \begin{figure}[!h]
           \includegraphics[width=0.98\textwidth]{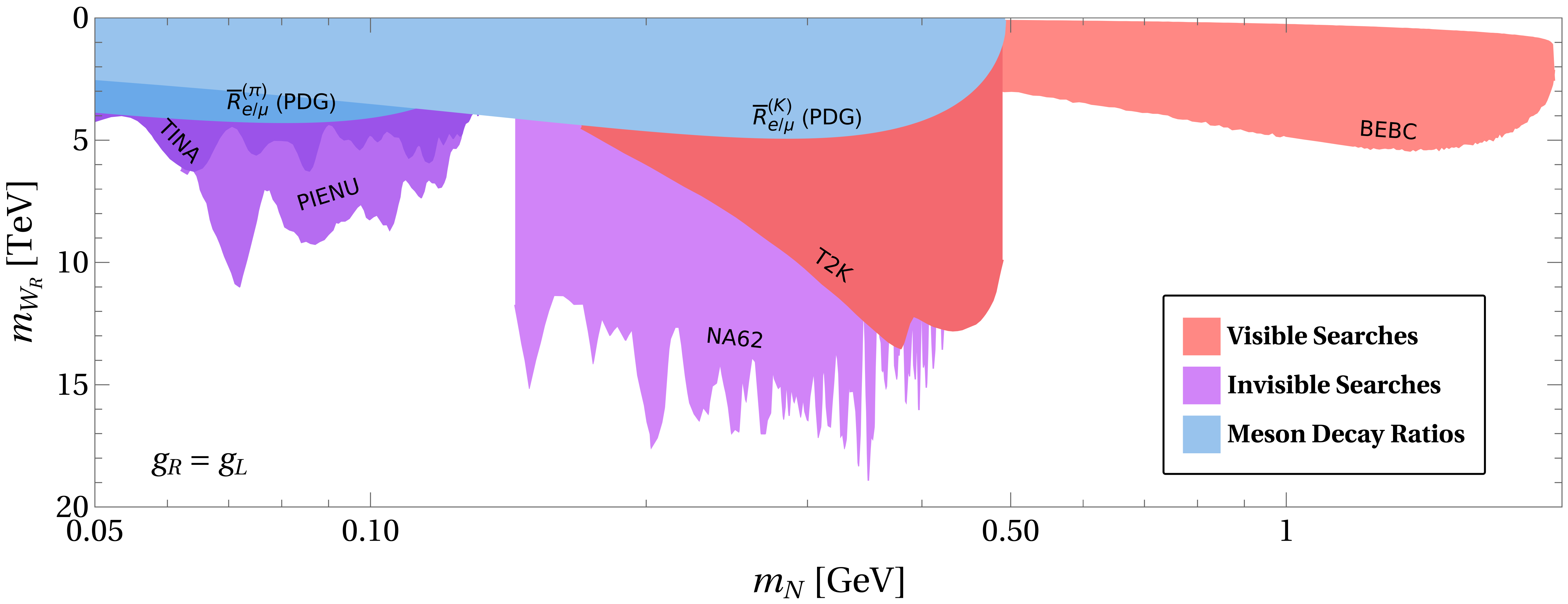}
        \caption{Bounds on  $m_W$ as a function of $m_N$ from visible searches (red) at
        T2K~\cite{T2K:2019jwa} and 
        BEBC~\cite{BEBCWA66:1986err}, 
        from  invisible peak 
        searches (purple) at PIENU~\cite{PIENU:2017wbj}, TINA~\cite{Britton:1992xv} and
        NA62~\cite{NA62:2020mcv}, as well as from $\pi$ and $K$ leptonic decay ratios (blue)~\cite{ParticleDataGroup:2022pth} at 90\% CL. We assume $\kappa=g_R/g_L=1$.}
            \label{fig:Bounds_MW}
        \end{figure}
    \end{center}
\end{widetext}

\textbf{\textit{Conclusions.}}---
Pseudoscalar meson decay experiments have been used in the past to set stringent limits on  active-sterile mixing. However,  it is conceivable that  this mixing  could be so tiny that it would be irrelevant for these decays. If right-handed currents exist, as predicted by LRSM, right-handed neutrinos can be produced in meson leptonic decays by a right-handed current, mediated by a vector boson $W_R$.
In this context, we have used low energy pseudoscalar meson leptonic decay data to constrain for the first time the mass $m_{W_R}$.

Our limits are valid for degenerate 
right-handed neutrinos with mass in the range $50 < m_N/{\rm MeV} < 2000$.
In this whole mass range they are at least as good as the LHC limits~\cite{ATLAS:2023cjo, CMS:2021dzb}, but in the region $60 \lesssim m_N/{\rm MeV} \lesssim 500$, they can be significantly more strict, specially due to NA62 and T2K,  we get 
$m_{W_R} \gtrsim (12-19)$ TeV at 90\% C.L.

Current experiments such as the ones in the Fermilab Short-Baseline Neutrino Program  which will collect data produced by typically $10^{21}$ POT  (ICARUS~\cite{ICARUS:2023gpo}, MicroBooNE~\cite{MicroBooNE:2016pwy}, SBND~\cite{MicroBooNE:2015bmn}) could  perhaps be used to improve these limits for $m_N<m_K$ using the conventional neutrino beam,  i.e. $\pi$ and $K$ leptonic decays. 
 Belle II~\cite{Belle-II:2022cgf} at SuperKEKB is 
 expected measure $\sim 10^{11}$ single $\tau$ decays. They may be able to  use $\tau \to \pi \, \nu_\tau$ to 
 probe $m_{W_R}$ up to $m_N < m_\tau-m_\pi$.
 The  future DUNE~\cite{DUNE:2020lwj,DUNE:2021tad} neutrino oscillation experiment
may also improve the bounds in 
 the region $m_N>m_K$ using production via prompt $D$ meson and $\tau$ decays.  The proposed HIKE (High-Intensity Kaon experiments)~\cite{HIKE:2022qra} at CERN could count with up to 6 times the NA62 beam intensity, being in 
 position, in principle, to increase significantly the sensitivity to $m_{W_R}$.  
We intend to investigate whether these experiments can  in fact do that in a future publication.

\vspace{-0.10cm}
\begin{acknowledgments}
R. Z. F. is partially supported by Fundação de Amparo à Pesquisa do Estado de São Paulo (FAPESP) under Contract No. 2019/04837-9, and Conselho Nacional de Desenvolvimento Científico e Tecnológico (CNPq).
C.S.F. acknowledges the support by FAPESP Contracts No. 2019/11197-6 and 2022/00404-3, and CNPq under Contract No. 407149/2021-0. 
L.P.S.L. and G.F.S.A. are fully financially supported by FAPESP under Contracts No. 2021/02283-6, 2022/10894-8 and No. 2020/08096-0. G.F.S.A. would like to thank the hospitality of 
the Fermilab Theory Group.
C. S. F. acknowledges support from the ICTP through the Associates Programme (2023-2028) while this work was being completed. We would like to thank P.A.N. Machado for careful reading and useful comments.
\end{acknowledgments}

\bibliography{T2KWR}
\end{document}